\documentstyle[12pt]{article}
\title{ Nonlinear Density Fluctuation Field Theory for
         Large Scale Structure }
\author{Yang Zhang \thanks{yzh@ustc.edu.cn} and Haixing  Miao\\
        Astrophysics Center \\
        University of Science and Technology of China \\
        Hefei, Anhui, China }
 \date{}

\begin{document}
\maketitle
\baselineskip=21truept

\newcommand{\be}{\begin{equation}}
\newcommand{\ee}{\end{equation}}
\newcommand{\ba}{\begin{eqnarray}}
\newcommand{\ea}{\end{eqnarray}}

\sf
\begin{center}
\Large  Abstract
\end{center}
\begin{quote}
 {\large
We develop the effective  field theory of density fluctuations for a
Newtonian self-gravitating $N$-body system in quasi-equilibrium,
apply it to  a homogeneous universe with small density fluctuations.
Keeping the density fluctuation up to the second order, we obtain
the nonlinear field equation of the 2-pt correlation  $\xi(r)$,
which contains the 3-pt correlation and formal ultra-violet
divergences. By the Groth-Peebles hierarchical ansatz and the mass
renormalization, the equation becomes closed with two new terms
beyond the Gaussian approximation, and their coefficients are taken
as parameters. The analytic solution is obtained in terms of the
hypergeometric functions, which is checked numerically. With one
single set of fixed two parameters, the correlation $\xi(r)$ and the
corresponding power spectrum $P(k)$ match simultaneously the results
from all the major surveys, such as APM, SDSS,  2dfGRS, and REFLEX.
The model gives a unifying  understanding of several seemingly
unrelated features of large scale structure from a field-theoretical
perspective. The theory is worthy to be extended to study the evolution
effects in an expanding universe.
}
\end{quote}

PACS numbers:


%\newpage
%\twocolumn
\baselineskip=21truept
\large

\section{INTRODUCTION}

Great progress has been made in understanding
the the large scale structure of the universe in past decades.
Not only observations of the major galaxy surveys
such as  SDSS  \cite{tegmark03,zehavi02,zehavi05},
2dF \cite{colless,hawkins,madgwick,percival},
APM \cite{maddox,padilla},
and REFLEX \cite{collins,schueker} etc,
have revealed the cosmic structures of increasingly large sky
dimension with new detailed features being found,
theoretical studies have also achieved important results,
through numerical simulations \cite{White,springel},
perturbation method \cite{peebles,davis,fry,goroff,bernardeau,valageas},
and thermodynamics \cite{saslaw}.
From  view point of dynamics, the
Universe filled with galaxies and clusters is a many-body
self-gravitating system in an asymptotic relaxed state,
since the cosmic time scale $1/H_0$ is longer than the
local crossing time scale \cite{saslaw}.
A systematic  approach to  statistical mechanics of
many-body systems is to convert the degrees of freedom of discrete
particles into a  continuous field.
Thereby, the fully-fledged techniques of field theory
can apply to study the systems  \cite{zustin}.
The Landau-Ginzburg theory is a known example in this regard.
We have formulated such a density field
theory of self-gravitating systems
and applied it to the large scales structure of the Universe
\cite{zhang}.
Under the Gaussian approximation,
the field equation of the 2-point correlation function and
the solution have been derived explicitly.
The result qualitatively interprets some observational features,
but it suffers from insufficient clustering on small scales.
In this paper, we will go beyond the Gaussian level and
include nonlinear terms of density fluctuations
up to the second order,
yielding a more satisfying description of the large scale structure.

\section{NONLINEAR FIELD EQUATION OF CORRELATION FUNCTION}

The universe is represented by a collection either
of galaxies, or of clusters, including their respective dark halos,
as the unit cells with random velocities.
Although the unit cell, galaxy or cluster, has different mass $m$,
both cases correspond to the same mass density $\rho(\bf r)$.
We study the asymptotic relaxed state of this Newtonian self-gravitating
system of $N$ points of mass $m$ with the Hamiltonian
$H = \sum_{i}^{N} \frac{p_i^2}{2m}
- \sum_{i<j}^N\frac{ Gm^2}{|\bf{r_i-r_j}|} $.
Thus the evolution effects will not be addressed in this paper.
By using the Hubbard-Stratonovich transformation \cite{zustin},
the grand partition function of this system at temperature $T$
can be cast into the generating functional as a path integral
$Z = \int D\phi\exp{[- \beta^{-1} \int d^3 {\bf{r}} {\cal L}(\phi)]}$,
where  $\phi$ is the gravitational field,
$ \beta \equiv 4\pi Gm/c_s^2$ of dimension $m^{-1}$,
$ c_s \equiv (T/m)^{1/2} $ the sound speed,
the effective Lagrangian is
${\cal L}(\phi)=\frac{1}{2}({\bigtriangledown} \phi)^2 -k_J^2 e^{\phi}$,
and $  k_J \equiv  (4\pi G\rho_0/c_s^2)^{1/2}$ is the Jeans wavenumber.
The term $-k_J^2 e^\phi $
has a minus sign because gravity is attractive.
By the Poisson's equation
$\bigtriangledown ^2 \phi ({\bf{r}}) + k_J^2 e^{\phi}=0$
the mass density $\rho$ is related to the $\phi$ field by
$\rho ({\bf{r}})  = mn({\bf{r}}) = \rho_0   e^{\phi(\bf r)}$.
So  $ \rho_0$ is the constant mass  density when $\phi =  0$.
We define  a  dimensionless re-scaled mass density field
$
\psi ({\bf{r}}) \equiv  e^{\phi ({\bf{r}})}
 =  \rho ({\bf{r}})/\rho_0,
$
and introduce an external source $J$
coupling with $\psi$ in the effective Lagrangian \cite{zhang}
\begin{equation}
{\cal L}(\psi, J)  = \frac{1}{2}
\left( \frac{\bigtriangledown \psi}{\psi} \right)^2 -k_J ^2  \psi -J\psi.
\end{equation}
 $J$ is used to handle the functional derivatives with ease.
So far $c_s$ is the only parameter in place of temperature $T$,
upon which $\beta$ and $k_J$ depend.
The field equation of $\psi$ in the presence of $J$ is
\be  \label{psij}
\bigtriangledown ^2 \psi -\frac{1}{\psi}(\bigtriangledown \psi)^2
+ k_J ^2 \psi^2 +J\psi^2 = 0 .
\ee
The $n$-point connected correlation function is given by
$ G^{(n)}_c ({\bf{r}}_1, \ldots , {\bf{r}}_n)
= \langle \delta\psi ({\bf{r}}_1 )\ldots
\delta\psi({\bf{r}}_n)\rangle$,
where
$ \delta\psi ({\bf{r}})= \psi( {\bf{r}}) -\langle\psi(\bf r)\rangle$
is the fluctuation field about the expectation value
$\langle\psi(\bf r) \rangle$,  and $\langle \delta \psi\rangle =0$.
A standard way to derive the field equation
of $G^{(2)} _c$ is to take functional derivative of
the expectation value of Eq.(\ref{psij})
w.r.t. the source $J$, and then set  $J=0$  \cite{goldenfeld,zhang}.
Assuming the large-scale homogeneity of the Universe
with a constant background density  $\langle \psi\rangle=\psi_0 $,
and keeping up to the second order of small fluctuation
\be \label{expansion}
\frac{1}{\psi} =  \frac{1}{\langle\psi \rangle+\delta\psi}
\simeq    \frac{1}{\langle\psi \rangle  }
\left(  1- \frac{\delta\psi}{\langle\psi \rangle }
+(\frac{\delta\psi}{\langle\psi \rangle})^2  \right),
\ee
we obtain the field equation of the 2-pt correlation function
\begin{eqnarray} \label{2peq}
&&\nabla^2G^{(2)}_c({\bf{r}}) +2k_J^2G^{(2)}_c({\bf{r}})  \nonumber\\
&+&\bigg[\frac{1}{\psi_0^2}G^{(2)}_c(\textbf{r})\nabla^2 G^{(2)}_c(0)
-\frac{1}{\psi_0} \nabla^2 G^{(3)}_c(0,\bf{r},\bf{r}) \nonumber\\
&&+\frac{2}{\psi_0 }\nabla G^{(2)}_c({\bf{r}})\cdot \nabla
G^{(2)}_c(0)  \bigg]
   =- \psi_0^2 \beta\delta^{(3)}(\textbf{r}),
\end{eqnarray}
where $ G^{(2)}_c (  {\bf{r}-\bf r'}) = G^{(2)}_c (  {\bf{r},\bf r'})=
\beta \delta   \langle \psi({\bf{r}})\rangle_{J=0}/\delta J({\bf{r'}})$
has been used.
Eq.(\ref{2peq}) is not closed, as it involves
the 3-pt correlation function  $ G^{(3)}_c$.
If higher order terms in $\delta \psi$ were allowed
in  Eq.(\ref{expansion}),
there would be $G^{(4)}_c$, etc,  in Eq.(\ref{2peq}).
Therefore, we have a hierarchy of field equations,
typical for the kinetic equation of a generic  many-body systems.
To close Eq.(\ref{2peq}),
we adopt the Groth-Peebles hierarchical ansatz \cite{groth-peebles}
\begin{eqnarray}\label{groth-peebles}
&&G^{(3)}_c({\bf r}_1,{\bf r}_2,{\bf r}_3)
=Q[G^{(2)}_c({\bf r}_1,{\bf r}_2)G^{(2)}_c({\bf r}_2,{\bf r}_3)+\nonumber\\
&&G^{(2)}_c({\bf r}_2,{\bf r}_3)G^{(2)}_c({\bf r}_3,{\bf r}_1)
 + G^{(2)}_c({\bf r}_3,{\bf r}_1)G^{(2)}_c({\bf r}_1,{\bf r}_2)],
\end{eqnarray}
where the constant $Q= 0.5\sim 1.0$ (\cite{fry,Efstathiou,Jing98,Jing04}).
Then,
Eq.(\ref{2peq}) becomes closed:
\begin{eqnarray} \label{eq2}
\nabla^2G^{(2)}_c({\bf r})
+ k^2_0 G^{(2)}_c ({\bf r})
+ {\bf a}\cdot \nabla G^{(2)}_c ({\bf r})
&-& b\Big(\nabla G^{(2)}_c ({\bf r}) \Big)^2 \nonumber\\
&=& -\psi^2_0\beta\delta^{(3)}({\bf r}),
\end{eqnarray}
where ${\bf a} \equiv \big(\frac{2}{\psi^2_0} -\frac{2Q}{\psi_0}\big)
\nabla  G^{(2)}_c(0)$,  $b \equiv Q/\psi_0>0$,
and
\be \label{renorm}
k^2_0\equiv 2 k_J^2 +(\frac{1}{\psi^2_0}-\frac{2Q}{\psi_0})
\nabla^2G^{(2)}_c(0).
\ee
Due to the higher order terms in Eq.(\ref{expansion}),
the friction term ${\bf a}\cdot \nabla G$  and
the nonlinear term $b(\nabla G)^2$  occur in Eq.(\ref{eq2}).
As is expected for an interacting field theory,
$ 2 k_J^2$ is modified by
an apparently divergent term $\nabla^2G^{(2)}_c(0)$.
We take $k_0^2$ as the physical wavenumber
like in the standard mass renormalization  \cite{zustin}.
Eq.(\ref{eq2}) is a nonlinear elliptic equation
with a point source $-\psi^2_0\beta\delta^{(3)}(\bf r)$.
Since $\psi^2_0\beta\propto m/c_s^2$, so
galaxies or clusters
with greater mass have a higher correlation amplitude.
This naturally explains
why the correlations of clusters or of galaxies
increase with richness and luminosity \cite{zhang}.
As mentioned earlier, galaxies  and clusters
are treated on equal footing  as  gravitating particles
differing only by  their masses,
their correlation functions have the same functional form,
differing only in the amplitude $\propto m/c_s^2$.
This is the observed facts \cite{guzzobartlett,bahcall-03}.
By isotropy of the Universe,
one puts $ G^{(2)}_c({\bf r})   \equiv \xi(r)$,
then Eq.(\ref{eq2}) reduces to
\be \label{eq3}
\xi''(x)+\Big(\frac{2}{x} +a\Big)\xi'(x)+\xi(x)-b \xi'\, ^2(x)
=-\psi_0^2\beta k_0 \frac{ \delta(x)}{x^2},
\ee
where  $x \equiv k_0r$, $\xi'\equiv \frac{d\xi}{dx}$,
and $a\equiv \big(\frac{2}{\psi^2_0}-\frac{2Q}{\psi_0}\big)\xi'(0)$.
Both $a$ and $b$ are treated as two parameters.

\section{ANALYTIC AND NUMERICAL SOLUTION}

The Gaussian approximation  \cite{chavanis,zhang}
is recovered by setting ${\bf a}=0=b$
and $k_0^2 \rightarrow 2k_J^2$ in Eq.(\ref{eq2}),
i.e. keeping  only the term
$\frac{1}{\psi} \simeq \frac{1}{ \langle\psi \rangle }$ in Eq.(\ref{expansion}).
The solution is $\xi(r) \propto A \frac{\cos (k_0r)}{ r}$
with $A= \frac{\psi_0^2 Gm}{c_s^2}$
and $k_0=(\frac{8\pi G\rho_0}{c_s^2})^{\frac{1}{2}}$,
and the power spectrum $P(k)=\frac{1}{2n}\frac{1}{(\frac{k}{k_0})^2-1}$,
where $n$ is the spatial number density.
This result qualitatively
explains several observed features,
such as a stronger correlation for more massive galaxies,
galaxies with a smaller  $n$ having a higher $ P(k) $
\cite{davis-geller,Einasto2002},
the scaling of ``correlation length'' $r_0$
with the mean cluster separation $d$
as $r_0(d) \propto d^{\,  0.3\sim 0.5}$
\cite{bahcall,bahcall-03,croft,Gonzalez,Zandivarez},
and damped oscillations of $\xi_c(r)$
for clusters with a wave-length $2\pi/k_0 \simeq 120 $ h$^{-1}$Mpc
\cite{broadhurst,Einasto1997a,Einasto1997b,Einasto2002,Tucker,Tago}.
Here one sees the physical meaning of the sound speed $c_s$.
Using the background mass density
$\rho_0=\rho_c\Omega_m =(3/8\pi G) H_0^2h^2\Omega_m$
leads to $c_s=\sqrt{3} H_0h\Omega_m^{1/2}/k_0$.
Taking the observed periodic length
$2\pi/k_0 \simeq 100\sim 120 h^{-1}$ Mpc
and $h\Omega_m^{1/2}\sim 0.36$ by WMAP \cite{spergel03,spergel07},
yields $c_s\simeq 1000 \sim 1200$km/s,
which is roughly that of
peculiar velocity of clusters or  galaxies.
Thus,  viewing $c_s$
as the random velocity of clusters or galaxies
is qualitatively consistent with
the observations of the periodicity in $\xi_c(r)$,
of $H_0$,  and of $h\Omega_m^{1/2}$.
The shortcomings of the Gaussian solution are  that
$\xi(r)$ is too low at small scales,
and that $P(k)$  has a sharp peak at $k=k_0$
and becomes negative for $k < k_0$ \cite{zhang}.

Now these problems are overcome by the nonlinear Eq.(\ref{eq3}),
whose solution is determined by the boundary condition
\be \label{bc}
 \xi(r_c)=C,\,\, \,\,\,\,\,\,  \xi'(r_c)=D,
\ee
at some $r_c$.
Note that, in fitting with observational data,
the amplitude $C$
is higher for clusters than for galaxies, as clarified earlier,
and the slope $D$ is roughly equal for clusters and for galaxies.
The range of parameters are taken
$a=(1.0\sim 1.3)$, $b = (0.01\sim 0.05)$,
$k_0 = (0.03\sim 0.06)  $ h\,Mpc$^{-1}$.
In computation we take $r_c \simeq 0.4/k_0$ h$^{-1}$Mpc.
Eq.(\ref{eq3}) is easily solved numerically.
In fact, it has an analytic solution as follows.
By perturbation method, since $b\ll 1$,
one sets the solution as a series
$\xi(x)=\sum^{\infty}_{i=0}b^i \xi_i(x)$.
Eq.(\ref{eq3}) becomes
$\ddot{\xi}_0+\Big(\frac{2}{x}+a\Big)\dot{\xi}_0+\xi_0
=-\psi_0^2\beta \frac{ \delta(r)}{x^2}$
and $\ddot{\xi}_i+\Big(\frac{2}{x}+a\Big)\dot{\xi}_i+\xi_i=g_i$,
where $g_i\equiv\sum^{i-1}_{j=0}\dot{\xi}_j\dot{\xi}_{i-1-j}$
for $  i>0$.
The solution for $\xi_0$ is
a linear combination of (\cite{gradshteyn-ryzhik})
\be \label{sol}
y_1\equiv e^{-\frac{1}{2}\alpha z}\Phi(\alpha, 2; z),\,\,\,
y_2\equiv e^{-\frac{1}{2}\alpha  z} \Psi(\alpha, 2; z),
\ee
where $\Phi$ and $\Psi$
are the degenerate hypergeometric functions,
$z\equiv (a^2-4)^{1/2}x$, and
$\alpha\equiv 1+\frac{a}{\sqrt{a^2-4}}$.
The solution of $\xi_i$ is a linear combination of Eq.(\ref{sol})
plus a particular solution
\be
 y_2(x) \int^{x}\frac{y_1(t)g_i(t)}{W(y_1,y_2)(t)}dt
-y_1(x) \int^{x}\frac{y_2(t)g_i(t)}{W(y_1,y_2)(t)}dt,
\ee
where $W(y_1,y_2)$ is the Wronskian.
The condition in Eq.(\ref{bc}) is satisfied by choosing
$\xi_0(r_c)=C$, $\xi_0'(r_c)=D$,
and $\xi_i(r_c)=\xi_i'(r_c)=0$ for $i>0$.
As is checked, up to the order $i=4$,
this analytic solution agrees with the numerical one.
Once $\xi(r)$ is known,
the power spectrum follows from Fourier transform
$P(k)=4\pi\int^{\infty}_0\xi(r)\frac{\sin(kr)}{kr}r^2dr$.
For galaxies from APM, 2dFGRAS and SDSS,
the calculated  $\xi_{gg}(r)$  and $P(k)$ are
shown in FIG.\ref{2pcf-of-all}, and in FIG.\ref{pk-of-all}, respectively.
For REFLEX X-ray clusters
the calculated $\xi_{cc}(r)$ and $P(k)$ are given
in FIG.\ref{REFLEXg3} and FIG.\ref{REFLEXp3}, respectively.
For SDSS clusters, $\xi_{cc}(r)$ is given in FIG.\ref{bahcall}.
It is seen that, for the fixed parameters
$(a,b) \simeq (1.2, 0.02)$, and  $k_0=0.05$ h Mpc$^{-1}$,
the calculated  $\xi_{gg}(r)$, $\xi_{cc}(r)$,
and  their respective $P(k)$
match very well all the major surveys for both galaxies and clusters,
simultaneously.
Thus, our density field theory
gives a decent description of the observational data.

\section{ DISCUSSION AND CONCLUSION}

Overall, the solution $\xi(r)$ of Eq.(\ref{eq3}) improves
the Gaussian one considerably, as shown FIG.\ref{landau-nonlinear},
in which the 2dFGRS active galaxies are taken for demonstration purpose.
It is found that the nonlinear term $-b(\xi') ^2$ has the effects
of strongly enhancing $\xi(r)$ on small scales,
and making  $P(k)$ flatter in $k<k_0$.
The friction term $a\xi'$ has the following effects:
slightly increasing the height of $\xi(r)$ on small scales;
moving the zeros of $\xi(r)$ to larger $r$;
strongly damping the amplitude of oscillations of $\xi$ on large scales
\cite{broadhurst,Tucker,Einasto1997a,Einasto1997b,Einasto2002,Tago},
as seen in FIG.\ref{oscillation};
smoothing out the sharp peak of $P(k)$ at $k_0$
and turning $P(k)$ positive for $k<k_0$.

The main conclusion of this paper
is the following.
The universe containing galaxies or clusters is viewed as
a many-particle system in asymptotic relaxed state
and can be described by an effective density field,
whereby the techniques of field theory applies,
yielding a perspective on the large scale structure of the universe
other than the conventional methods.
There appears  the Jeans scale
$k_0\simeq (0.04\sim 0.06)$ hMpc$^{-1}$,
which is the unique scale underlying the large scale structure
as a gravitational system.
Up to nonlinear terms $(\delta \psi)^2$
beyond the Gaussian approximation,
the nonlinear field equation of the 2-pt correlation function
of the density fluctuations has been derived
and solved analytically.
This analytic result of field theory interprets
several observed features of large scale structures.
With fixed values of the parameters $(a,b, k_0)$,
the solution matches the observed $\xi(r)$ and $P(k)$
of both galaxies and of clusters, simultaneously.

Although our results match  the observational data
on large scales very well,
our model is still preliminary  at the present stage,
there are several problems that need to be further
addressed in the future as in the following.

Firstly the calculated $\xi_{gg}(r)$ for galaxies
increases too fast on very small scales $r\leq 0.2 h^{-1}$Mpc.
This indicates that
the model may break down on such small scales close to a galaxy size.
This may suggest that
either higher order terms of perturbations should be taken into account,
or the effects of galaxy formation
and local virilization need to be included.

Next,  there is a limitation in applying the Groth-Peebles ansatz
in Eq.(\ref{groth-peebles}).
As is known,
for descriptions of any many-particle dynamic system based on Gibbs-Boltzmann equation,
a procedure is usually taken, which decomposes the complicated equation
into a set of differential equations,
so that each one in the set is possibly manageable.
However, thereby,
a hierarchy of BBKGY type, or the like, is inevitably arises.
Different treatments of the hierarchy
are employed for different systems,
and a cut-off of the hierarchy is usually is used.
For instance, in the case of the photon gas of CMB,
the multipole decomposition is involved
for the temperature anisotropies and polarization,
and the common practice is to cut off the hierarchy of multipoles by
letting higher multipole components to zero,
yielding a very accurate description of CMB spectra
\cite{ZhaoZhang,Zhang et al,XiaZhang}).
But for the calculation of $G^{(2)}_c({\bf r})$ in our context,
one can not set $G^{(3)}_c({\bf r}_1,{\bf r}_2,{\bf r}_3)$  to zero,
since these are important and give rise to nonlinear effects,
due to the  long range nature of gravity force.
The Groth-Peebles ansatz has been used in our analytic treatment,
because it is a simple one and  an analytic solution can be derived.
We would like to mention that the ansatz only approximately reflects
the actual distribution
since the forms of factor, $Q$,  is in fact a function of $r$.
In this case Eq.(\ref{eq2}) for $G^{(2)}_c({\bf r})$
would be more complicated
and  analytic solutions
for the general case would be difficult to derive explicitly.

Thirdly, in fitting the observation of galaxies,
we have not separated the dark matter from galaxies in our present model.
Therefore, no bias is introduced
and the baryon acoustic oscillations are not
incorporated in our model.
A comprehensive treatment of two components, dark matter and baryons,
in our theory would require substantial extensions to the model
discussed the present paper.

Finally, it should be mentioned that
our theoretical model deals with only the quasi-relaxed state
of the large scale structure of the universe,
which, as an assumption, is qualitatively good approximation
since the overall expansion rate $H_0$ is smaller than
the particle collision rate.
It would be much desired to have an extension of the present model
to take into account of an evolutionary description.
These issues  will to be addressed in our further studies.

ACKNOWLEDGMENT: Y. Zhang's research work was supported by the CNSF
No.10773009, SRFDP, and CAS.

%\newpage

\newpage

\begin{figure}
\centerline{%\includegraphics[width=12cm]{fig1.eps}
}
\caption{\label{2pcf-of-all}   $\xi_{gg}(r)$
fits the galaxy correlation functions of the surveys,
APM, SDSS, and 2dFGRS, simultaneously.}
\end{figure}

\begin{figure}
\centerline{%\includegraphics[width=12cm]{fig2.eps}
}
\caption{ $P(k)$ of FIG.\ref{2pcf-of-all} also
fits the power spectra of APM, SDSS, and 2dFGRS.
$P(k)\propto k^{-1.5}$ in $(0.05\sim 0.5)$hMpc$^{-1}$.\label{pk-of-all}}
\end{figure}

\begin{figure}
\centerline{%\includegraphics[width=12cm]{fig3.eps}
}
\caption{\label{REFLEXg3}
$\xi_{cc}(r)$ fits the correlation
function of REFLEX X-ray clusters (\cite{collins}). }
\end{figure}

\begin{figure}
\centerline{%\includegraphics[width=12cm]{fig4.eps}
}
\caption{\label{REFLEXp3}
$P(k)$ from FIG.\ref{REFLEXg3} fits the corresponding
spectrum of REFLEX X-ray clusters (\cite{schueker}). }
\end{figure}

\begin{figure}
\centerline{%\includegraphics[width=12cm]{fig5.eps}
}
\caption{\label{bahcall}
$\xi_{cc}(r)$ of SDSS clusters of different
richness (\cite{bahcall-03}) are obtained by the same set $(a,b)$
but different initial amplitude. }
\end{figure}

\begin{figure}
\centerline{%\includegraphics[width=12cm]{fig6.eps}
}
\caption{\label{landau-nonlinear}
The solution $\xi_{gg}(r)$ of the nonlinear Eq.(\ref{eq3})
improves the Gaussian one on small scales
and matches 2dFGRAS active galaxies.}
\end{figure}

\begin{figure}
\centerline{%\includegraphics[width=12cm]{fig7.eps}
}
\caption{\label{oscillation} The calculated $\xi_{cc}(r)$
with oscillations compared with the observed of X-ray clusters
(\cite{Einasto2002,Tago}).}
\end{figure}

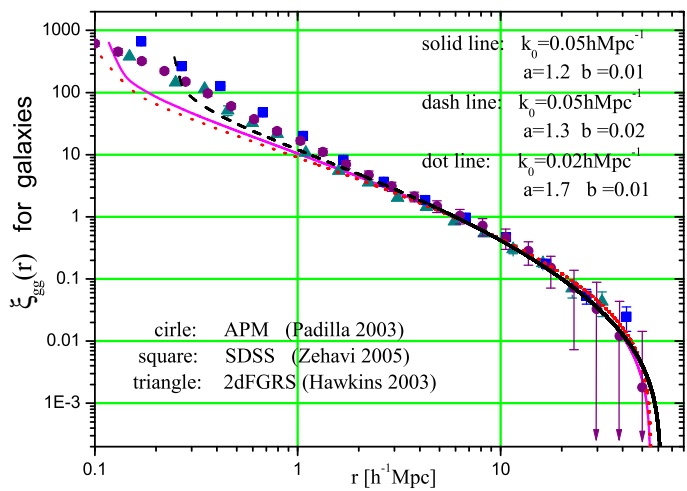
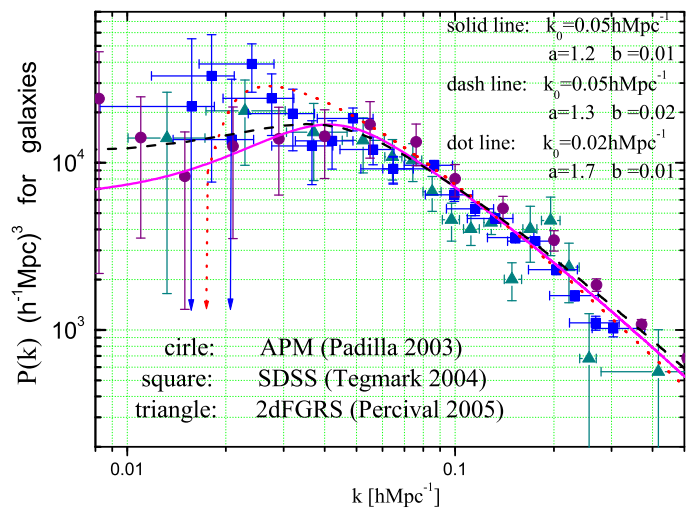
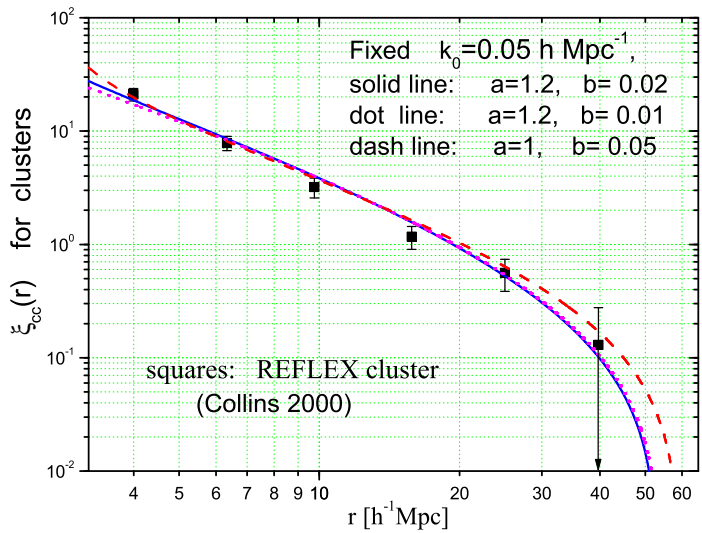
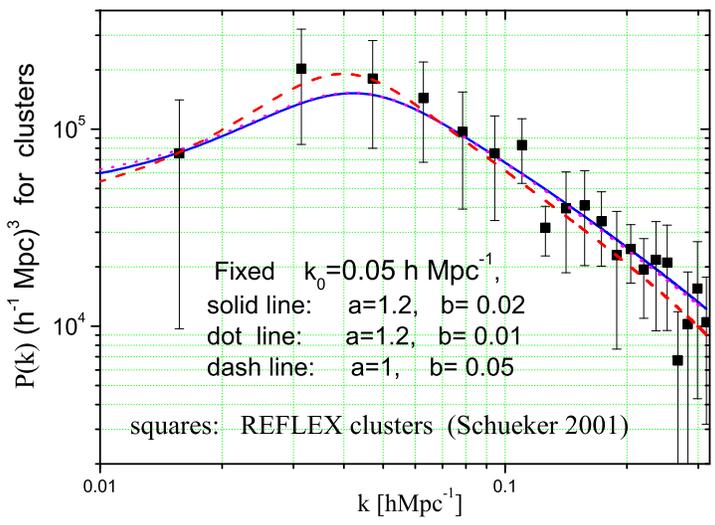
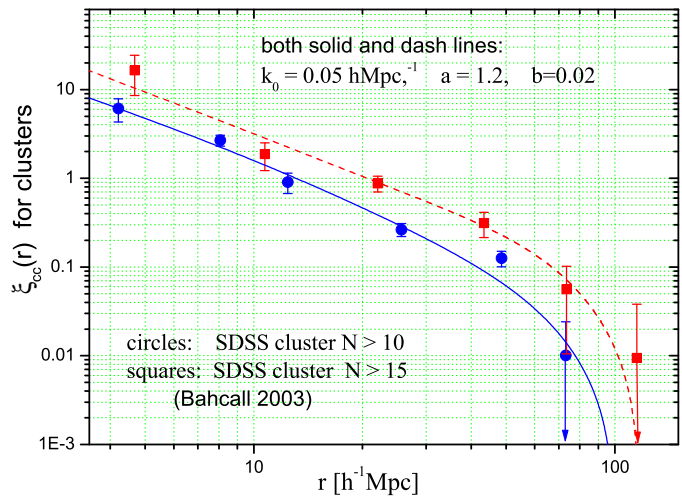
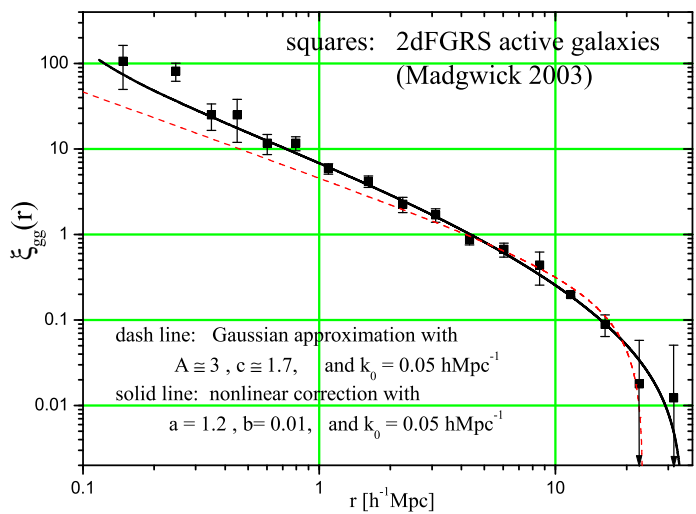
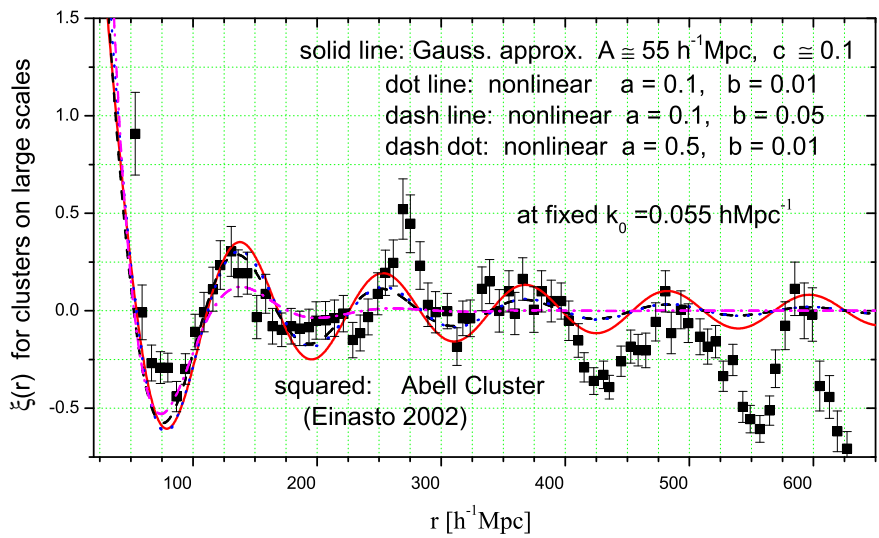
\end{document}